\pdfoutput=1

\documentclass[runningheads,a4paper]{llncs}

\usepackage{graphicx}

\usepackage[numbers,sort&compress,sectionbib]{natbib}
\usepackage{amsmath}
\usepackage{algorithm}
\usepackage{algorithmic}
\usepackage{array}
\usepackage{multirow}
\usepackage[caption=false]{subfig}

\usepackage{amssymb}
\usepackage{url}

\DeclareMathOperator{\Normal}{\mathcal{N}}

\newcommand{\R}{\mathbb{R}}
\newcommand{\given}{\,|\,}

\urldef{\mails}\path|{xukevin,hero}@umich.edu|
\newcommand{\keywords}[1]{\par\addvspace\baselineskip
\noindent\keywordname\enspace\ignorespaces#1}

\begin{document}

\mainmatter  

\title{Dynamic stochastic blockmodels: \\
	Statistical models for time-evolving networks}

\titlerunning{Dynamic stochastic blockmodels}

%
%
\author{Kevin S.~Xu%
\thanks{Current affiliation: 3M Corporate Research Laboratory, 
St.~Paul, MN, USA}%
\and Alfred O. Hero III}
%

\institute{Department of Electrical Engineering and Computer Science,\\
University of Michigan, Ann Arbor, MI, USA\\
\mails}

%
%

\maketitle

\begin{abstract}
Significant efforts have gone into the development of statistical models for 
analyzing data in the form of networks, such as social networks. 
Most existing work has focused on modeling static networks, which represent 
either a single time snapshot or an aggregate view over time. 
There has been recent interest in statistical modeling of 
\emph{dynamic networks}, 
which are observed at multiple points in time and offer a richer 
representation of many complex phenomena. 
In this paper, we propose a state-space model for dynamic networks that 
extends the well-known \emph{stochastic blockmodel} for static networks 
to the dynamic setting. 
We then propose a procedure to fit the model using a modification of the 
extended Kalman filter augmented with a local search. 
We apply the procedure to analyze a dynamic social network of email 
communication. 
\keywords{dynamic network, stochastic blockmodel, state-space model}
\end{abstract}

\section{Introduction}
Many complex physical, biological, and social phenomena are naturally 
represented by networks. 
Tremendous efforts have been dedicated to analyzing network data, which 
has led to the development of many formal statistical models for networks. 
Most research has focused on static networks, which 
either represent a single time snapshot of the phenomenon  
being investigated or an aggregate view over time. 
As such, statistical models for static networks have a long history in 
statistics and sociology among other fields \citep{Goldenberg2010}. 
However, most complex phenomena, including social behavior, 
are time-varying, which has led researchers to consider dynamic, 
time-evolving networks.

In this paper, we consider dynamic networks represented 
by a sequence of snapshots of the network at discrete time steps. 
We characterize such networks using a set of unobserved \emph{time-varying 
states} from which the observed snapshots are derived. 
We propose a state-space model for dynamic networks that combines two types of 
statistical models: a static model for the individual 
snapshots and a temporal model for the evolution of the states. 
The network snapshots are modeled using the stochastic blockmodel 
\citep{Holland1983}, a simple parametric model commonly used in the analysis 
of static social networks. 
The state evolution is modeled by a stochastic dynamic system. 
Using a Central Limit Theorem approximation, we develop a near-optimal 
procedure for fitting the proposed model in the on-line setting where only 
past and present network snapshots are available. 
The inference procedure involves a modification of the extended Kalman 
filter, which is used for state tracking in many applications 
\citep{Haykin2001}, augmented with a local search strategy. 
We apply the proposed procedure to analyze 
a dynamic social network of email communication and predict future email 
activity. 

\section{Related work}
\label{sec:Related_work}
Several statistical models for dynamic networks have previously been 
proposed by extending a static model to the dynamic setting in a similar 
fashion to our proposed model \citep{Goldenberg2010}.
Two such models include temporal extensions 
of the exponential random graph model \citep{AhmedPNAS2009} and latent space 
model \citep{WestveldAOAS2011}. 
More closely related to the state-space model we propose are several temporal 
extensions of stochastic blockmodels (SBMs). 
SBMs divide nodes in the network 
into multiple classes and generate edges independently with probabilities 
$\theta_{ab}$ dependent on the class memberships $a,b$ of the nodes 
\citep{Holland1983}. 
\citet{YangML2011} propose a dynamic SBM involving a transition matrix that 
specifies the probability that a node in class $i$ at time $t$ switches to 
class $j$ at time $t+1$ for all $i,j,t$ and fit the model using 
Gibbs sampling and simulated annealing. 
\citet{HoAISTATS2011} propose a temporal extension of a mixed-membership 
version of the SBM using linear state-space models for the class 
membership vectors of node clusters. 
One major difference between \citep{YangML2011,HoAISTATS2011} and this paper 
is that we treat the edge probabilities $\theta_{ab}$ as 
\emph{time-varying states}, 
while \citep{YangML2011,HoAISTATS2011} treat them as time-invariant 
parameters. 
In addition, our model allows for a simpler inference procedure using a 
Central Limit Theorem approximation. 
We demonstrate the importance of the time-varying states for analysis of a 
dynamic social network in Section \ref{sec:Enron}. 

\section{Static stochastic blockmodels}
\label{sec:Static_SBM}
We first introduce notation and summarize the static 
stochastic blockmodel (SSBM), which we use 
as the static model for the individual network snapshots. 
We represent a dynamic network by a time-indexed sequence of graphs, with 
$W^t = [w_{ij}^t]$ denoting the adjacency matrix of the graph 
observed at time step $t$. 
$w_{ij}^t = 1$ if there is an edge from node $i$ to node $j$ at time $t$, 
and $w_{ij}^t = 0$ otherwise. 
We assume that the graphs are directed, 
i.e.~ $w_{ij}^t \neq w_{ji}^t$ in general, and that 
there are no self-edges, i.e.~$w_{ii}^t = 0$. 
$W^{(s)}$ denotes the set of all snapshots up to time $s$, $\{W^s, 
W^{s-1}, \ldots, W^1\}$. 
The notation $i \in a$ indicates that node $i$ is a member of class $a$. 
$|a|$ denotes the number of nodes in class  $a$. 
The classes of all nodes at time $t$ is given by a vector 
$\vec{c}^t$ with $c_i^t = a$ if $i \in a$ at time $t$. 
We denote the submatrix of $W^t$ corresponding to the relations between 
nodes in class $a$ and class $b$ by $W_{[a][b]}^t$. 
We denote the vectorized equivalent of a matrix $X$, i.e.~the vector 
obtained by simply stacking columns of $X$ on top of one another, by 
$\vec{x}$. 
Doubly-indexed subscripts such as $x_{ij}$ denote entries of matrix $X$, 
while singly-indexed subscripts such as $x_i$ denote entries of the 
vectorized equivalent $\vec{x}$. 

Consider a snapshot at an arbitrary time step $t$. 
An SSBM is parameterized by a $k \times k$ matrix $\Theta^t = [\theta_{ab}^t]$, 
where $\theta_{ab}^t$ denotes the probability of forming an edge between a 
node in 
class $a$ and a node in class $b$, and $k$ denotes the number of classes. 
The SSBM decomposes the adjacency matrix into $k^2$ blocks, where each block 
is associated with relations between nodes in two classes $a$ and $b$. 
Each block corresponds to a submatrix $W_{[a][b]}^t$ of the adjacency 
matrix $W^t$.
Thus, given the class membership vector $\vec{c}^t$, each entry of $W^t$ is 
an independent realization of a Bernoulli random variable with a 
block-dependent parameter; that is, $w_{ij}^t \sim \text{Bernoulli} 
\Big(\theta_{c_i^t c_j^t}^t\Big)$. 

SBMs are used in two settings:
\begin{enumerate}
\item The \emph{a priori} blockmodeling setting, 
where class memberships are known or assumed, and the objective is to 
estimate the matrix of \emph{edge probabilities} $\Theta^t$.

\item The \emph{a posteriori} blockmodeling setting, where the objective 
is to simultaneously estimate $\Theta^t$ and the class membership 
vector $\vec{c}^t$.
\end{enumerate}
Since each entry of $W^t$ is independent, the likelihood for the SBM is 
given by
\begin{align}
	f\left(W^t; \Phi^t\right) &= \prod_{i \neq j} \left( 
		\theta_{c_i c_j}^t\right)^{w_{ij}^t} \left(1 - 
		\theta_{c_i c_j}^t\right)^{1-w_{ij}^t} \nonumber \\
	\label{eq:Likelihood_Phi_simp}
	&= \exp\left\{\sum_{a=1}^k \sum_{b=1}^k 
		\left[m_{ab}^t \log \left(\theta_{ab}^t\right) + \left(n_{ab}^t - 
		m_{ab}^t\right) \log \left(1 - \theta_{ab}^t\right)\right]\right\},
\end{align}
where $m_{ab}^t = \sum_{i \in a} \sum_{j \in b} w_{ij}^t$ denotes the number 
of \emph{observed} edges in block $(a,b)$, and 
\begin{equation}
	\label{eq:N_def}
	n_{ab}^t = 
	\begin{cases}
		|a||b| & a \neq b \\
		|a|(|a|-1) & a = b
	\end{cases}
\end{equation}
denotes the number of \emph{possible} edges in block $(a,b)$ 
\citep{Karrer2011}.
The parameters are given by $\Phi^t = \Theta^t$ in the a priori setting, and 
$\Phi^t = \{\Theta^t,\vec{c}^t\}$ in the a posteriori setting. 
In the a priori setting, a sufficient statistic for estimating $\Theta^t$ is 
the matrix $Y^t$ of \emph{block densities} (ratio of observed 
edges to possible edges within a block) with entries 
$y_{ab}^t = m_{ab}^t / n_{ab}^t$.
$Y^t$ also happens to be the maximum-likelihood estimate of $\Theta^t$, 
which can be shown \citep{Karrer2011} by setting the derivative of the 
logarithm of \eqref{eq:Likelihood_Phi_simp} to $0$. 

Estimation in the a posteriori setting is more involved, and 
many methods have been proposed, including Gibbs sampling 
\citep{NowickiJASA2001}, 
label-switching \citep{Karrer2011,Zhao2012}, and spectral clustering 
\citep{Sussman2012}. 
The label-switching methods use a heuristic for solving the combinatorial 
optimization problem of maximizing the likelihood 
\eqref{eq:Likelihood_Phi_simp} 
over the set of possible class memberships, which is too large to 
perform an exhaustive search. 

\section{Dynamic stochastic blockmodels}
We propose a state-space model for dynamic networks that consists of a temporal 
extension of the static stochastic blockmodel. 
First we present the model and inference procedure for a priori blockmodeling, 
and then we discuss the additional steps necessary for a posteriori 
blockmodeling. 
The inference procedure is on-line, i.e.~the state estimate at time $t$ is 
formed using only observations from time $t$ and earlier.

\subsection{A priori blockmodels}
\label{sec:Inference_known}
In the a priori SSBM setting, $Y^t$ is a sufficient statistic for 
estimating $\Theta^t$ as discussed in Section \ref{sec:Static_SBM}. 
Thus in the a priori dynamic SBM setting, we can equivalently treat $Y^t$ as 
the observation rather than $W^t$. 
The entries of $W_{[a][b]}^t$ are independent and identically distributed 
(iid) $\text{Bernoulli}\left(\theta_{ab}^t\right)$; 
thus by the Central Limit Theorem, the sample mean $y_{ab}^t$ is approximately 
Gaussian with mean $\theta_{ab}^t$ and variance
$(\sigma_{ab}^t)^2 = \theta_{ab}^t(1-\theta_{ab}^t) / n_{ab}^t$,
where $n_{ab}^t$ was defined in \eqref{eq:N_def}. 
We assume that $y_{ab}^t$ is indeed Gaussian for all $(a,b)$ and posit 
the linear observation model 
\begin{equation*}
	Y^t = \Theta^t + Z^t,
\end{equation*}
where $Z^t$ is a zero-mean iid
Gaussian noise matrix with variance $(\sigma_{ab}^t)^2$ for the 
$(a,b)$th entry. 

\begin{figure}[t]
	\centering
	\includegraphics[width=4.5in]{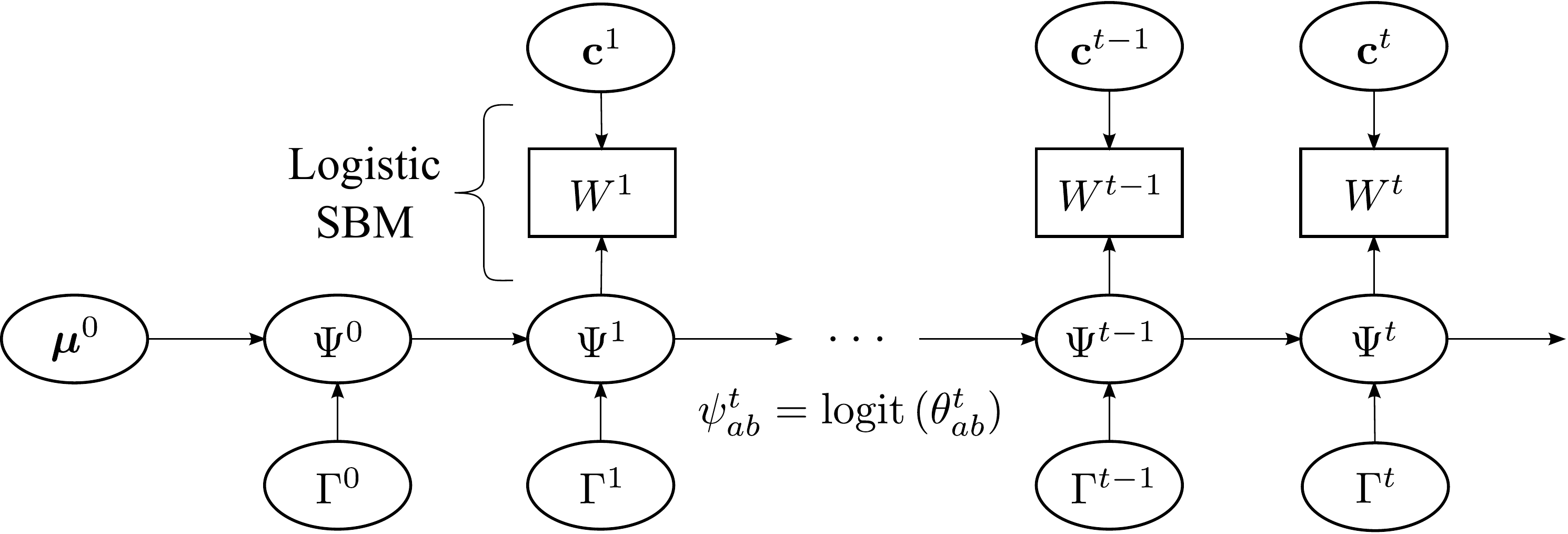}
	\caption[Graphical representation of the proposed model] 
		{Graphical representation of the proposed model. 
		The rectangular boxes denote observed quantities, and the ovals 
		denote unobserved quantities.
		The logistic SBM refers to applying the logistic function to each 
		entry of $\Psi^t$ to obtain $\Theta^t$ then generating $W^t$ using 
		$\Theta^t$ and $\vec{c}^t$.}
	\label{fig:Graphical_model_proposed}
\end{figure}

In the dynamic setting where past snapshots are 
available, the observations would be given by the set $Y^{(t)}$. 
The set $\Theta^{(t)}$ can then be viewed as states of a dynamic 
system that is generating the noisy observation sequence. 
We complete the model by specifying a model for the state evolution over time. 
Since $\theta_{ab}^t$ is a probability and must be bounded between $0$ and $1$, 
we instead work with the matrix $\Psi^t = [\psi_{ab}^t]$ where $\psi_{ab}^t = 
\log(\theta_{ab}^t) - \log(1 - \theta_{ab}^t)$, the logit of $\theta_{ab}^t$. 
A simple model for the state evolution is the random walk 
\begin{equation*}
	\vec{\psi}^t = \vec{\psi}^{t-1} + \vec{v}^t,
\end{equation*}
where $\vec{\psi}^t$ is the vector representation of the matrix $\Psi^t$, and 
$\vec{v}^t$ is a random vector of zero-mean Gaussian entries, 
commonly referred to as process noise, with covariance matrix $\Gamma^t$. 
The entries of the process noise vector are not necessarily independent or 
identically distributed (unlike the entries of $Z^t$) to allow for states 
to evolve in a correlated manner. 
The observation model can then be written 
in terms of $\vec{\psi}^t$ as\footnote{Note that we have converted the block 
densities $Y^t$ and observation noise $Z^t$ to their respective vector 
representations $\vec{y}^t$ and $\vec{z}^t$.}
\begin{equation}
	\label{eq:Obs_model_SBM}
	\vec{y}^t = h\left(\vec{\psi}^t\right) + \vec{z}^t,
\end{equation}
where the function $h: \R^{k^2} \rightarrow \R^{k^2}$ is defined by 
$h_i(\vec{x}) = 1/(1+e^{-x_i})$, i.e.~the 
logistic function applied to each entry of $\vec{x}$. 
We denote the covariance matrix of $\vec{z}^t$ by $\Sigma^t$, which is a 
diagonal 
matrix\footnote{The indices $(a,b)$ 
for $(\sigma_{ab}^t)^2$ are converted into a single index $i$ 
corresponding to the vector representation $\vec{z}^t$.} with entries given by 
$(\sigma_{ab}^t)^2$. 
A graphical representation of the proposed model for the dynamic network 
is shown in Fig.~\ref{fig:Graphical_model_proposed}. 

To perform inference on this model, we assume the initial state is Gaussian 
distributed, 
i.e.~$\vec{\psi}^0 \sim \Normal\left(\vec{\mu}^0, \Gamma^0\right)$, and that 
$\{\vec{\psi}^0, \vec{v}^1, 
\ldots, \vec{v}^t, \vec{z}^1, \ldots, \vec{z}^t\}$ are mutually independent. 
If \eqref{eq:Obs_model_SBM} was linear in $\vec{\psi}^t$, then the optimal 
estimate of $\vec{\psi}^t$ in terms of minimum mean-squared error 
would be given by the Kalman filter \citep{Haykin2001}. 
Due to the non-linearity, we apply the extended Kalman filter (EKF), which 
linearizes the dynamics about the predicted state and provides an 
near-optimal estimate of $\vec{\psi}^t$. 
The predicted state under the random walk model is simply 
$\vec{\hat{\psi}}^{t|t-1} = \vec{\hat{\psi}}^{t-1|t-1}$ with covariance 
$R^{t|t-1} = R^{t-1|t-1} + \Gamma^t$. 
Let $J^t$ denote the Jacobian of $h$ evaluated at the predicted state 
$\vec{\hat{\psi}}^{t|t-1}$. 
The EKF update equations are as follows \citep{Haykin2001}:
\begin{align*}
	&\text{Near-optimal Kalman gain:} && K^t = R^{t|t-1} \left(J^t\right)^T 
		\left[J^t R^{t|t-1} \left(J^t\right)^T + \Sigma^t\right]^{-1} \\
	&\text{Posterior state estimate:} && \vec{\hat{\psi}}^{t|t} 
		= \vec{\hat{\psi}}^{t|t-1} + K^t\left[\vec{y}^t 
		- h\left(\vec{\hat{\psi}}^{t|t-1}\right)\right] \\
	&\text{Posterior estimate covariance:} && R^{t|t} = \left(I 
		- K^t J^t\right) R^{t|t-1}
\end{align*}
The posterior state estimate $\vec{\hat{\psi}}^{t|t}$ provides a near-optimal  
fit to the model at time $t$ given the observed sequence $W^{(t)}$. 
How to choose the hyperparameters $\left(\vec{\mu}^0,\Gamma^0,\Sigma^t, 
\Gamma^t\right)$ in an optimal manner is beyond the scope of this paper and is 
discussed in \citep[chap.~5]{Xu2012a}. 

\subsection{A posteriori blockmodels}
\label{sec:Inference_unknown}
In many applications, the class memberships $\vec{c}^t$ are not known a 
priori and must be estimated along with $\Psi^t$. 
This can be done using label-switching methods 
\citep{Karrer2011,Zhao2012}, but rather than maximizing 
the likelihood, we 
maximize the posterior state density given the entire sequence of 
observations $W^{(t)}$ up to time $t$ to account for the prior information. 
This is done by alternating between label-switching and applying the EKF. 

The posterior state density is given by
\begin{equation}
	\label{eq:Psi_posterior}
	f\left(\vec{\psi}^t \given W^{(t)}\right) \propto f\left(W^t 
		\given \vec{\psi}^t, W^{(t-1)}\right) f\left(\vec{\psi}^t \given 
		W^{(t-1)}\right).
\end{equation}
By the conditional independence of current and past observations given the 
current state, $W^{(t-1)}$ drops out of the first term in 
\eqref{eq:Psi_posterior}. 
It can thus be obtained simply by substituting $h(\vec{\psi}^t)$ 
for $\vec{\theta}^t$ in \eqref{eq:Likelihood_Phi_simp}. 
The second term in \eqref{eq:Psi_posterior} is equivalent to 
$f\left(\vec{\psi}^t \given \vec{y}^{(t-1)}\right)$ because the class 
memberships at all previous time steps have already been estimated. 
By applying the Kalman filter to the linearized temporal model 
\citep{Haykin2001}, $f\left(\vec{\psi}^t \given \vec{y}^{(t-1)}\right) \sim 
\Normal\big(\vec{\hat{\psi}}^{t|t-1}, R^{t|t-1}\big)$. 
Thus the logarithm of the posterior density is given by
\begin{equation}
	\label{eq:Psi_posterior_simp}
	\begin{split}
	\log f\Big(\vec{\psi}^t \given &W^{(t)}\Big) = 
		c - \frac{1}{2} \left(\vec{\psi}^t - \vec{\hat{\psi}}^{t|t-1}\right)^T
		\left(R^{t|t-1}\right)^{-1} \left(\vec{\psi}^t 
		- \vec{\hat{\psi}}^{t|t-1}\right) \\
	&+\sum_{a=1}^k \sum_{b=1}^k \left\{m_{ab}^t 
		\log \left[h\left(\psi_{ab}^t\right)\right] + \left(n_{ab}^t - 
		m_{ab}^t\right) \log \left[1 - h\left(\psi_{ab}^t\right)\right]\right\},
	\end{split}
\end{equation}
where $c$ is a constant term independent of $\vec{\psi}^t$ that can be 
ignored\footnote{At the 
initial time step, $\vec{\hat{\psi}}^{1|0} = \vec{\mu}^0$ and 
$R^{1|0} = \Gamma^0+\Gamma^1$.}. 

We use the log-posterior \eqref{eq:Psi_posterior_simp} 
as the objective function for label-switching. 
We find that a simple local search (hill climbing) algorithm 
\citep{Russell2003} initialized using the 
estimated class memberships at the previous time step suffices, because 
only a small fraction of nodes change classes between time steps in most 
applications. 
At the initial time step, we employ the spectral clustering algorithm of 
\citet{Sussman2012} for the SSBM as the initialization. 

\section{Application to Enron email network}
\label{sec:Enron}
We demonstrate the proposed procedure on a dynamic social network constructed 
from the Enron corpus \citep{PriebeCMOT2005,Priebe2009}, which 
consists of about $0.5$ million email messages between $184$ Enron employees 
from 1998 to 2002. 
We place directed edges between employees $i$ and 
$j$ at time $t$ if $i$ sends at least one email to $j$ during week $t$. 
Each time step corresponds to a $1$-week interval. 
We make no distinction between emails sent ``to'', ``cc'', or ``bcc''. 
In addition to the email data, the roles of most of the employees within the 
company (e.g.~CEO, president, manager, etc.) are available, 
which we use as classes for a priori blockmodeling. 
Employees with unknown roles are placed in an ``others'' class. 

\subsection{State tracking}

\begin{figure}[tp]
	\centering
	\subfloat[Week $59$: a normal week]{\includegraphics[width=2.3in] 
		{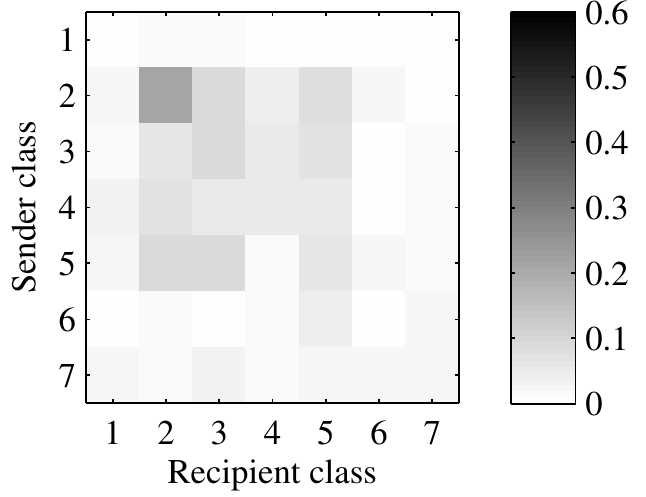}}
	\quad
	\subfloat[Week $89$: CEO Skilling resigns] 
		{\includegraphics[width=2.3in]{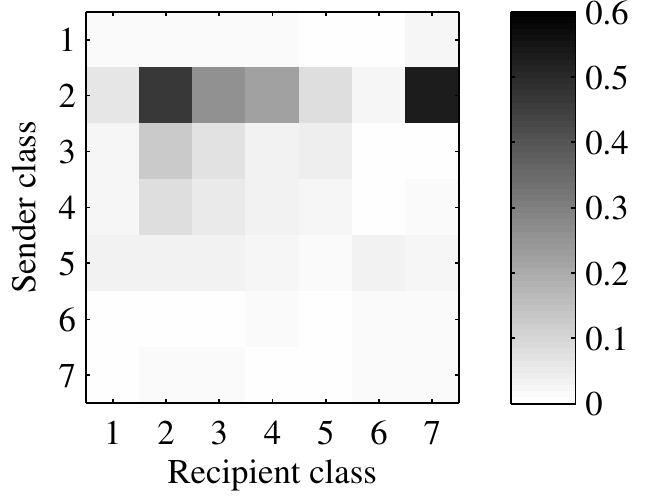}}
	\caption[Estimated edge probability matrices for two selected weeks]
		{Estimated edge probability matrices for two selected weeks. 
		Entry $(i,j)$ denotes the estimated probability of an edge from 
		class $i$ to class $j$. 
		Classes are as follows: (1) directors, (2) CEOs, (3) presidents, 
		(4) vice-presidents, (5) managers, (6) traders, and (7) others. 
		Notice the increase in the probability of edges from CEOs during the 
		week of Skilling's resignation.}
	\label{fig:Enron_heatmaps}
\end{figure}

\begin{figure}[tp]
	\centering
	\subfloat[Presidents to presidents]
		{\includegraphics[width=4.6in]{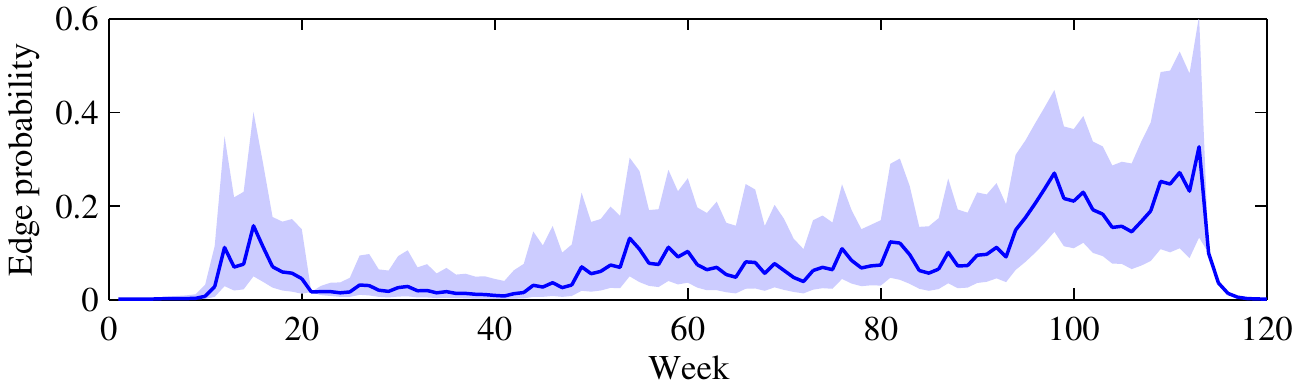}} \\
	\subfloat[Others to others]
		{\includegraphics[width=4.6in]{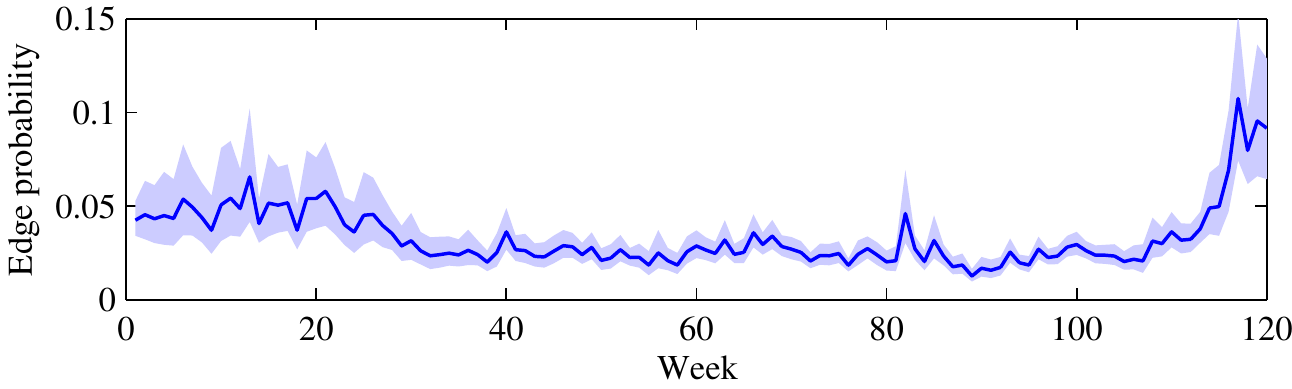}}
	\caption[A priori EKF estimated edge probabilities 
		$\hat{\theta}_{ab}^{t|t}$ 
		with $95\%$ confidence intervals for selected $a,b$ by week]
		{A priori EKF estimated edge probabilities $\hat{\theta}_{ab}^{t|t}$ 
		(solid lines)
		with $95\%$ confidence intervals (shaded region) for selected 
		$a,b$ by week. 
		An increase in edge probabilities between Enron presidents (a) occurs 
		prior to a similar increase between those in other roles (b) 
		suggesting insider knowledge.}
		\label{fig:Enron_from_CEO}
\end{figure}

We begin by examining the temporal variation of the states, which we refer to 
as \emph{state tracking}. 
Recall that the states $\Psi^t$ correspond to the logit of the edge 
probabilities $\Theta^t$. 
We first apply the a priori EKF to obtain the state estimates 
$\vec{\hat{\psi}}^{t|t}$ and their variances (the diagonal of $R^{t|t}$). 
Applying the logistic function, 
we can then obtain the estimated edge probabilities $\hat{\Theta}^{t|t}$ 
with confidence intervals. 

Examining the temporal variation of $\hat{\Theta}^{t|t}$ 
reveals some interesting trends. 
For example, a large increase in the probabilities of edges from CEOs is 
found at week $89$. 
This is the week in which CEO Jeffrey Skilling resigned and is 
confirmed to be the cause of the increased probabilities by examining 
the content of the emails. 
Fig.~\ref{fig:Enron_heatmaps} shows a comparison of the matrix 
$\hat{\Theta}^{t|t}$ during a normal week and during the week Skilling 
resigned. 

Another interesting trend is highlighted in Fig.~\ref{fig:Enron_from_CEO}, 
where the temporal variation of two selected edge probabilities 
over the entire data trace with $95\%$ confidence intervals is shown. 
Edge probabilities between Enron presidents show a 
steady increase as Enron's financial situation worsens, hinting at more 
frequent and widespread insider discussions, while emails between others 
(not of one of the six known roles) begin to increase only after Enron 
falls under federal investigation. 

A key observation from this analysis is the importance of modeling the 
edge probabilities as time-varying states, as opposed to time-invariant 
parameters as in \citep{YangML2011,HoAISTATS2011}. 
Indeed the \emph{temporal variation} of the edge probabilities is what 
reveals the internal dynamics of this time-evolving social network. 
Furthermore, the temporal model provides estimates with less uncertainty 
than the static SBM, with $95\%$ confidence intervals that are $24\%$ 
narrower on average.

\subsection{Dynamic link prediction}

Next we turn to the task of dynamic link prediction, 
which differs from static link 
prediction \citep{Liben-Nowell2007} because the link predictor must 
simultaneously predict the new edges that will be formed at time $t+1$, as 
well as the current edges (as of time $t$) that will disappear at time 
$t+1$, from the observations $W^{(t)}$. 
The latter task is not addressed by most static link prediction methods in the 
literature. 

\begin{figure}[t]
	\centering
	\includegraphics[width=2.8in]{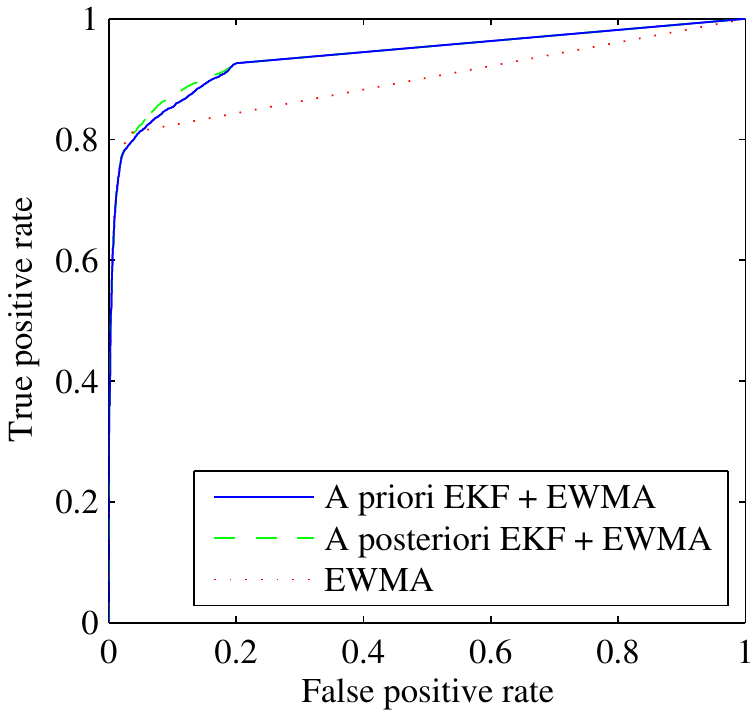}
	\caption[Comparison of ROC curves for link prediction on Enron data] 
		{Comparison of ROC curves for link prediction on Enron data. 
		True positive rate denotes the fraction of actual edges that are 
		correctly predicted, and false positive rate denotes the fraction of 
		non-edges that are predicted to be edges. The convex combination of 
		either EKF with the EWMA outperforms the EWMA alone by accounting for 
		block-level characteristics.}
	\label{fig:Enron_ROC}
\end{figure}

Since the SBM assumes stochastic equivalence between nodes in the same class, 
the EKF alone is only a good predictor of the block densities $Y^t$, not the 
edges themselves. 
However, the EKF can be combined with a predictor that operates 
on individual edges to form a link predictor. 
A simple individual-level predictor is the exponentially-weighted 
moving average (EWMA) given by $\hat{W}^{t+1} = \lambda \hat{W}^{t} + 
(1-\lambda) W^t$. 
Using a convex combination of the EKF and EWMA predictors, we obtain 
a better link predictor that incorporates both 
block-level characteristics (through the EKF) and individual-level 
characteristics (through the EWMA). 
This can be seen from the receiver operating characteristic (ROC) curves 
in Fig.~\ref{fig:Enron_ROC}. 
The a posteriori EKF slightly outperforms the a priori EKF 
because the a posteriori EKF finds a better 
fit to the dynamic SBM via a better assignment of nodes to classes 
than the a priori (assumed) assignment. 

\section{Conclusion}
This paper proposes a statistical model for dynamic networks that utilizes 
a set of unobserved time-varying states to characterize the dynamics of 
the network. 
The proposed model extends the well-known stochastic blockmodel for static 
networks to the dynamic setting can be used for either a priori or a 
posteriori blockmodeling. 
The main contribution of the paper is a near-optimal on-line inference 
procedure for the proposed model using a modification of the 
extended Kalman filter, augmented with a local search. 
We applied the proposed inference procedure to the Enron email network and 
discovered some interesting trends when we examined the estimated states. 
One such trend was a steady increase in emails between Enron presidents 
as Enron's financial situation worsened, while emails between other employees 
remained at their baseline levels until Enron fell under federal 
investigation. 
In addition, the proposed procedure showed promising results for predicting 
future email activity. 
We believe the proposed model and inference procedure can be applied to 
reveal the internal dynamics of many other dynamic networks.

\subsubsection*{Acknowledgments.}
This work was partially supported by the 
Army Research Office grant W911NF-12-1-0443. 
Kevin Xu was partially supported by an award from the Natural 
Sciences and Engineering Research Council of Canada.

\bibliographystyle{splncsnat}
\bibliography{abrv,references,library_sa}

\begin{thebibliography}{16}
\providecommand{\natexlab}[1]{#1}
\providecommand{\url}[1]{\texttt{#1}}
\providecommand{\urlprefix}{}

\bibitem[{Ahmed and Xing(2009)}]{AhmedPNAS2009}
Ahmed, A., Xing, E.P.: Recovering time-varying networks of dependencies in
  social and biological studies.
\newblock Proc. Nat. Acad. Sci. 106(29), 11878--11883 (2009)

\bibitem[{Goldenberg et~al.(2010)Goldenberg, Zheng, Fienberg, and
  Airoldi}]{Goldenberg2010}
Goldenberg, A., Zheng, A.X., Fienberg, S.E., Airoldi, E.M.: A survey of
  statistical network models.
\newblock Found. Trends Mach. Learn. 2(2), 129--233 (2010)

\bibitem[{Haykin(2001)}]{Haykin2001}
Haykin, S.: {Kalman filtering and neural networks}.
\newblock Wiley-Interscience (2001)

\bibitem[{Ho et~al.(2011)Ho, Song, and Xing}]{HoAISTATS2011}
Ho, Q., Song, L., Xing, E.P.: Evolving cluster mixed-membership blockmodel for
  time-varying networks.
\newblock In: Proceedings of the 14th Int. Conf. Artif. Intell. Statist. (2011)

\bibitem[{Holland et~al.(1983)Holland, Laskey, and Leinhardt}]{Holland1983}
Holland, P.W., Laskey, K.B., Leinhardt, S.: {Stochastic blockmodels: First
  steps}.
\newblock Soc. Netw. 5(2), 109--137 (1983)

\bibitem[{Karrer and Newman(2011)}]{Karrer2011}
Karrer, B., Newman, M.E.J.: Stochastic blockmodels and community structure in
  networks.
\newblock Phys. Rev. E 83, 016107 (2011)

\bibitem[{Liben-Nowell and Kleinberg(2007)}]{Liben-Nowell2007}
Liben-Nowell, D., Kleinberg, J.: {The link-prediction problem for social
  networks}.
\newblock J. Am. Soc. Inf. Sci. 58(7), 1019--1031 (2007)

\bibitem[{Nowicki and Snijders(2001)}]{NowickiJASA2001}
Nowicki, K., Snijders, T.A.B.: Estimation and prediction for stochastic
  blockstructures.
\newblock J. Am. Stat. Assoc. 96(455), 1077--1087 (2001)

\bibitem[{Priebe et~al.(2005)Priebe, Conroy, Marchette, and
  Park}]{PriebeCMOT2005}
Priebe, C.E., Conroy, J.M., Marchette, D.J., Park, Y.: Scan statistics on
  {E}nron graphs.
\newblock Comput. Math. Organ. Theory 11(3), 229--247 (2005)

\bibitem[{Priebe et~al.(2009)Priebe, Conroy, Marchette, and Park}]{Priebe2009}
Priebe, C.E., Conroy, J.M., Marchette, D.J., Park, Y.: {Scan statistics on
  Enron graphs} (2009),
  \urlprefix\url{http://cis.jhu.edu/~parky/Enron/enron.html}

\bibitem[{Russell and Norvig(2003)}]{Russell2003}
Russell, S.J., Norvig, P.: {Artificial intelligence: A modern approach}.
\newblock Prentice Hall, 2nd edn. (2003)

\bibitem[{Sussman et~al.(2012)Sussman, Tang, Fishkind, and
  Priebe}]{Sussman2012}
Sussman, D.L., Tang, M., Fishkind, D.E., Priebe, C.E.: {A consistent adjacency
  spectral embedding for stochastic blockmodel graphs}.
\newblock arXiv:1108.2228v3 [stat.ML]  (2012)

\bibitem[{Westveld and Hoff(2011)}]{WestveldAOAS2011}
Westveld, A.H., Hoff, P.D.: A mixed effects model for longitudinal relational
  and network data, with applications to international trade and conflict.
\newblock Ann. Appl. Stat. 5(2A), 843--872 (2011)

\bibitem[{Xu(2012)}]{Xu2012a}
Xu, K.S.: {Computational methods for learning and inference on dynamic
  networks}.
\newblock Ph.D. thesis, University of Michigan (2012)

\bibitem[{Yang et~al.(2011)Yang, Chi, Zhu, Gong, and Jin}]{YangML2011}
Yang, T., Chi, Y., Zhu, S., Gong, Y., Jin, R.: Detecting communities and their
  evolutions in dynamic social networks---a {B}ayesian approach.
\newblock Mach. Learn. 82(2), 157--189 (2011)

\bibitem[{Zhao et~al.(2012)Zhao, Levina, and Zhu}]{Zhao2012}
Zhao, Y., Levina, E., Zhu, J.: {Consistency of community detection in networks
  under degree-corrected stochastic block models}.
\newblock The Annals of Statistics (in press) (2012)

\end{thebibliography}

\end{document}